\documentclass[fleqn,10pt]{wlscirep}
\usepackage[utf8]{inputenc}
\usepackage[T1]{fontenc}

\usepackage[english]{babel}
\usepackage{amsmath}
\usepackage{amsfonts}
\usepackage{mathtools}

\usepackage{graphicx}
\usepackage{hyperref}
\usepackage{color,soul}
\usepackage{float}

\title{Loss of sustainability in scientific work}

\author[1]{Niklas Reisz}
\author[1]{Vito D.P. Servedio}
\author[1,2,3]{Vittorio Loreto}
\author[1]{William Schueller}
\author[1]{Márcia R. Ferreira}
\author[1,4,5,*]{Stefan Thurner}
\affil[1]{Complexity Science Hub Vienna, Josefst\"adter Strasse 39, A-1080 Vienna, Austria}
\affil[2]{Sony Computer Science Lab, 6, Rue Amyot, 75005, Paris, France}
\affil[3]{Sapienza Unversity of Rome, Physics Dept, Piazzale A. Moro 2, 00185, Rome, Italy}
\affil[4]{Section for Science of Complex Systems, CeMSIIS, Medical University of Vienna, Spitalgasse 23, A-1090, Austria}
\affil[5]{Santa Fe Institute, 1399 Hyde Park Road, Santa Fe, NM 85701, USA}
\affil[*]{stefan.thurner@meduniwien.ac.at}

\begin{abstract}
For decades the number of scientific publications has been rapidly increasing, effectively out-dating knowledge at a tremendous rate. Only few scientific milestones remain relevant and continuously attract citations. Here we quantify how long scientific work remains being utilized, how long it takes before today's work is forgotten, and how milestone papers differ from those forgotten. To answer these questions, we study the complete temporal citation network of all American Physical Society journals. We quantify the probability of attracting citations for individual publications based on age and the number of citations they have received in the past. We capture both aspects, the forgetting and the tendency to cite already popular works, in a microscopic generative model for the dynamics of scientific citation networks. We find that the probability of citing a specific paper declines with age as a power law with an exponent of $\boldsymbol{\alpha \sim -1.4}$. Whenever a paper in its early years can be characterized by a scaling exponent above a critical value, $\boldsymbol{\alpha_c}$, the paper is likely to become "ever-lasting". We validate the model with out-of-sample predictions, with an accuracy of up to 90\% (AUC $\boldsymbol{\sim 0.9}$). The model also allows us to estimate an expected citation landscape of the future, predicting that 95\% of papers cited in 2050 have yet to be published. The exponential growth of articles,  combined with a power-law type of forgetting and papers receiving fewer and fewer citations on average, suggests a worrying tendency toward information overload and raises concerns about scientific publishing's long-term sustainability.
\end{abstract}
\begin{document}

\flushbottom
\maketitle

\thispagestyle{empty}

\section*{Introduction}

When asked by a student, "Dr. Einstein, aren't these the same questions as last year's [physics] final exam?", he replied: "Yes; But this year the answers are different." To Einstein, it was very clear that his field was evolving at a rapid pace and that new scientific discoveries outdated historical knowledge. This observation might be even more relevant today. Since Einstein's days, the global scientific output, measured in the number of publications per year, has grown exponentially. The number of articles published every year in all of physics  increased a thousandfold from 200 papers in 1900 to about 200,000 in 2010~\cite{Sinatra2015x, Martin2013}. The total scientific output is about ten times higher than in physics \cite{NSF}.

Scientific work usually builds on pre-existing knowledge, which is acknowledged through citations in publications. Only a small fraction of this knowledge is actively accessed and actually cited \cite{Meho_2007}. We find that less than half of all papers published in American Physics Society (APS) journals received more than one citation during the last ten years. Most works become outdated quickly, are never looked at, are forgotten and  never mentioned again. This is especially true for the increasing number of non-reproducible papers that do not contribute knowledge sustainably~\cite{Loken2017, Ioannidis2005}. On the other hand, there are scientific milestone papers -- typically exceptional works -- that continue to attract citations, even after decades~\cite{Redner2005}. What emerges from the process of knowledge creation, selection, and forgetting is a ``living'' or active body of knowledge  that is actively used -- ideally -- for pushing the scientific frontier forward. The active knowledge is a tiny subset of all the existing knowledge stored in papers.
Its extent can be estimated using citation data. %Citations also play a central role to determine the success of scientific works and their authors \cite{Noorden2010}. 

Large-scale publication and citation data has become increasingly available in recent years. It opens up the possibility to reconstruct the history of science (production and reception) on the granular basis of individual publications. In particular, it becomes possible to trace citations in so-called citation networks.

% I MOVED THE SECTION about citation networks from here to the methods section
There are immediate lessons that have been learned from citation networks that are important in our context. Early contributions date back to the 1960s \cite{deSollaPrice1965} where citation networks were pioneered by investigating links between scientific publications. Along with other insights, most strikingly it was noted, that the percentage of papers receiving $n$ citations in any year approximately scales as a power law in citations $n$. It follows that most papers receive a small number of citations, while a few papers in the ''fat tail'' of the distribution receive many. This observation has been confirmed~\cite{seglen1992, Redner1998}. In follow-up work, an explanation was offered in terms of ''cumulative advantage''~\cite{Price1976}, stating that the probability for a paper to get cited is proportional to the number of citations it accumulated up to that point in time. This process is also called \textit{preferential attachment}~\cite{Barabsi1999x}. It has been studied in depth and confirmed by a growing number of authors, both in a theoretical context and in the context of citation networks~\cite{Krapivsky2001, Dorogovtsev2002, Albert2002, Newman2003-x, Newman2001, Capocci2006, Jeong_2003}.

While preferential attachment is a plausible mechanism for understanding the structure of citation networks, recent studies suggest that preferential attachment alone does not result in accurate predictions of several features of citation networks~\cite{Borner2004, Lehmann2005}. Other factors need to be taken into account. Timing of publication has been found to be a contributing factor and a first-mover advantage was proposed along with preferential attachment~\cite{Newman2009x}. If a work is among the first in an emerging field, then there is not much competition, and it may receive more citations. It was noted that papers that receive citations unusually quickly tend to continue to receive more citations than usual, even though they might not be the most cited papers at the time~\cite{Newman_2014x}. 

Time not only plays a role in the success of a paper, but also in its ``decline''. In this context, the notion of ``half-life of literature'' has been introduced some time ago~\cite{Burton1960}. It is defined as the time during which the second half of all papers that are still being cited were published. This time is rather short, because most papers cite the recent past~\cite{Burton1960}. This fact was also observed in~\cite{deSollaPrice1965, Redner2005}. In contrast to the  half-life of nuclei, the ``half-life of literature'' is not a constant, but depends on the overall growth of the citation network. For that reason, citation networks must always be considered against the exponential growth of the number of publications. In physics, this growth is characterized by a doubling time of about 11.8 years~\cite{Martin2013}. In ~\cite{Martin2013, Fanelli2016} it is noted that the number of publications per author (productivity) has practically not increased during the last century while the average number of authors per paper has increased~\cite{Martin2013}. The observed growth in the number of publications can therefore be largely attributed to the growth in the overall number of scientists and research funding. From the existence of a half-life of literature and the growth of the network follows, immediately, that the distribution of the age of citations (time between the publication of the citing and the cited paper) is highly skewed towards more recent papers. %~\cite{Redner2005}.
This general trend also persists if one accounts for the exponential growth of the number of articles published each year~\cite{Redner2005}. For this reason it is apparent, that there is a strong preference of papers to cite more recent works, an effect that is often referred to as \textit{aging}~\cite{Wang2013x, Yin2017}.

Preferential attachment in combination with aging are the two mechanisms that form the basis for a number of mathematical models that reasonably explain citation networks and their dynamics. In~\cite{Dorogovtsev2000}, an evolving network model with preferential attachment and a power law aging of nodes is considered. It demonstrates a strong first mover advantage. Another model with preferential attachment and power-law aging~\cite{Safdari2016} shows that if aging is included in the model, the associated degree distribution of the resulting network deviates from a strict power law. They confirm their findings on a Hollywood actor network. In~\cite{Borner2004} an author-paper network is simulated, where authors can read, write, and cite papers, and search for references in a recursive manner. Aging is included by fitting a Weibull distribution to the time-dependent attachment probability of papers. This model explains the deviations from strict power laws in the degree distribution and is supported by empirical data from PNAS articles. A stochastic model of citation dynamics that builds on~\cite{Borner2004} combines preferential attachment and aging with a recursive search and copy mechanism~\cite{Golosovsky2017}. The main result is a non-linearity in the network growth that allows for individual papers to achieve diverging citation trajectories.  The rate at which both, patents and scientific articles, acquire citations is studied in~\cite{Higham2017, Higham2017_2}. There, this rate is modeled with power law preferential attachment and exponential aging. Strikingly, in this model, the aging component is completely independent of the preferential attachment.  Yet another model for patent citation networks assumes the probability to cite a patent to follow linear preferential attachment, as well as power law forgetting with an exponential cutoff~\cite{Valverde2007}. It presents a method to determine this probability from empirical data and shows that the scaling behavior can be explained by the combination of preferential attachment and aging. In all the mentioned models, preferential attachment and aging have an impact on the fate of a paper. Some models allow for short-term performance predictions (citations)~\cite{Newman_2014x, Golosovsky2017, Higham2017_2}; some also do  predictions on longer horizons~\cite{Wang2013x} of individual papers. 

What has been studied in much less detail is the dynamics of scientific work being forgotten. For how long is scientific work relevant?  How long will it take before today’s work is forgotten? Which works have the potential to become milestones, and under what conditions does this happen? Can these conditions be identified by seeing papers not in isolation, but as a part of the citation network? 
What has been missing in this context so far is a model that focuses on the forgetting aspect in the context of a growing citation network. We want to add that missing piece to understand the dynamics of forgetting in order to better estimate the future state of the system as a whole. From that, we can then understand how certain papers stay relevant for  decades, while most papers are forgotten.

\begin{figure}[h!]
	\centering
	  \includegraphics[width=0.7\columnwidth]{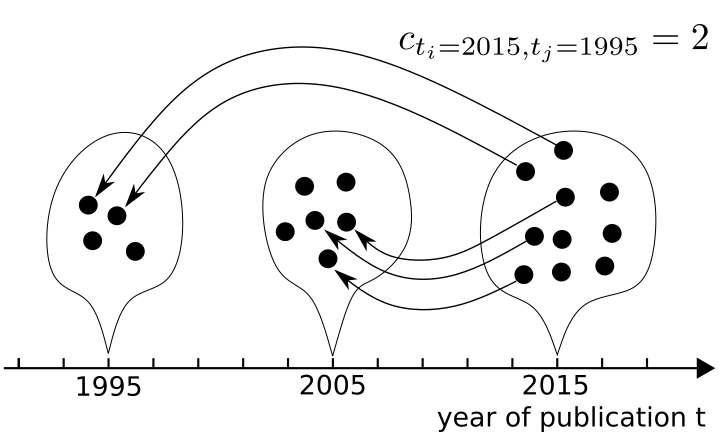}\\
		\caption{ 
		Number of citations, $c_{t_i,t_j}$, originating from papers published in year $t_i$ and citing papers published in year $t_j$.}
	\label{fig:aor}
\end{figure}

%{\bf A framework for forgetting.}
To understand the forgetting process more quantitatively, we introduce a variable to represent the total number of citations originating from papers published in year $i$ that cite papers published in year $j$,
\begin{equation}
c_{t_i t_j} \equiv \sum_{k,l} 
\begin{cases}
M_{kl}, & \text{ if publication year of $k$} = t_j \\
& \text{ and publication year of $l$} = t_i,\\
0 & \text{ else}
\end{cases};
\end{equation}
This is shown in Fig.~\ref{fig:aor}, where each dot represents a paper published in a particular year and each arrow represents a citation from one paper to another.

\begin{figure}[h!]
	\centering
	  \includegraphics[width=0.7\columnwidth]{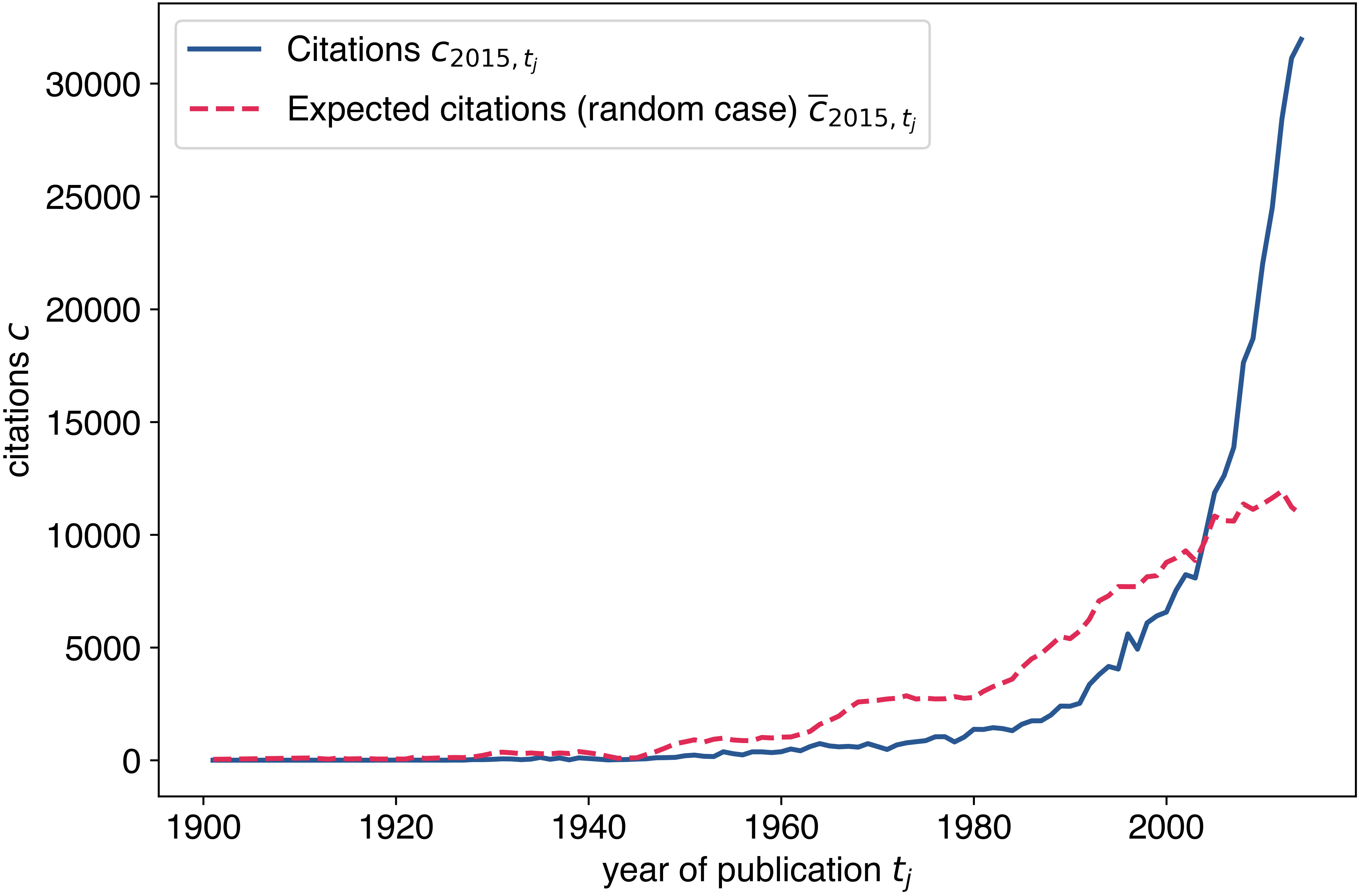}
	  \includegraphics[width=0.68\columnwidth]{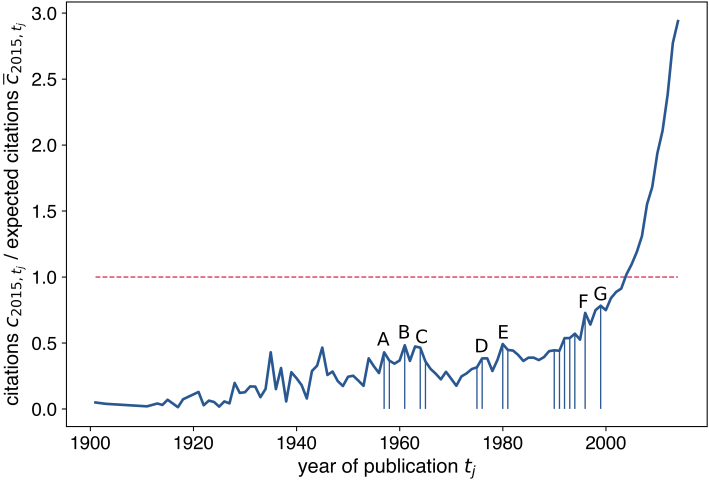}
		\caption{ 
		(a) Number of citations per year from papers published in 2015 (blue solid line) and expected values of the random case, where citations are assigned to publications randomly with equal probability (red dashed line). The red line is proportional to the number of papers published each year.
		(b) ``Forgetting curve'' or citation preference of papers published in 2015. Values $>1$ indicate a relative preference with respect to equal probability citations, values $<1$ indicate an under-representation.
		Vertical lines mark the publishing years of milestone scientific papers such as:
		A: "Theory of Superconductivity" (Bardeen, Cooper,Schrieffer); B: "Effects of Configuration Interaction on Intensities and Phase Shifts" (Fano); C: "Inhomogeneous Electron Gas" (Hohenberg, Kohn); D: "Special points for Brillouin-zone integrations" (Monkhorst, Pack); E: "Ground State of the Electron Gas by a Stochastic Method"
(Ceperley, Alder); F: "Generalized Gradient Approximation Made Simple" (Perdew, Burke, Ernzerhof) \& "Efficient iterative schemes for ab initio total-energy calculations using a plane-wave basis set" (Kresse, Furthmüller); G: "From ultrasoft pseudopotentials to the projector augmented-wave method" (Kresse, Joubert)}
	\label{fig:aor2}
\end{figure}

In Fig.~\ref{fig:aor2} (a) the values of $c_{t_i=2015,t_j}$ (citations per year from papers published in 2015) are shown (blue) for all American Physical Society journals that include about 600,000 papers published between 1893 and 2015 and around 5,000,000 citations. Next, we look at a hypothetical situation, where all papers have the same probability of being cited, independent of their age. We call the corresponding variable $\overline{c}_{t_it_j}$ (red dashed line). We can see that $\overline{c}_{t_it_j}$ is directly proportional to the number of papers published in every year and can be used as a normalization factor, to account for the growing number of publications. By dividing $c_{t_i=2015,t_j} / \overline{c}_{t_i=2015,t_j}$, we get a measure that is independent of the number of papers published each year.  We call this resulting curve the ``forgetting curve'' of the APS, shown in Fig.~\ref{fig:aor2} (b). Note a general trend of a strong preference towards citing papers that are not older than ten years. In a randomized scenario, where new publications would cite other publications with equal probability, the forgetting curve would be a constant at 1. Note that spikes in the forgetting curve coincide with the publishing years of milestone papers. These heavily cited papers stand out from the bulk of papers and cause the corresponding year to be over-represented. We find that in years, where a milestone paper was published, these papers, while making up only $0.03\%$ of the papers published, attract on average $14\%$ of all citations to papers published in that year. For more details, see~\hyperref[SI1]{SI 1}.

The forgetting curve can be used to infer the age structure of the literature new knowledge is built on. We explain the curve with the help of a simple model of citation dynamics that takes into account in different ways the bulk of  publications and the exceptional works.

%{\bf Model for citation dynamics.}
To model the dynamical process of citing literature, we devise a simple generative model. We consider the case where a new paper $i$ is published at the time $t_i$ and ask how likely will the paper $i$ cite the paper $j$? To answer it, we measure the function that modulates the  probability of attracting citations for individual publications. This function is based on the age of the referenced paper $j$, $t = t_i - t_j$, and the number of citations the paper $j$ received in the past, up to time $t_i$. We call this function the {\em attachment kernel},  $f(k,t)$. Assuming that the underlying mechanisms of how we choose our references do not change over time, the dependence of $t_i$ and $t_j$ of the attachment kernel can be condensed into one variable $t$. To measure the attachment kernel, we adapt a method used in \cite{Newman2001} and apply it to the 120 years of empirical citation data of the APS; see~\hyperref[SI2]{SI 2}. The resulting kernel can be approximated well by a linear function in $k$ and a power law decay in time,
\begin{equation}
   f(k, t)= (k+1)(t+1)^{-\alpha} \quad.
\end{equation}
The exponent $\alpha$ cannot be completely decoupled from the number of citations $k$, as it shows a slight dependency $\alpha = \alpha(k) $. However, we show that for the bulk of papers, $\alpha$ can be well approximated as a constant from the empirical data by its mean value over all observations. For the mean, we find a value of $\overline{\alpha} = 1.42$.  We confirm the linear preferential attachment in $k$ found in other works \cite{Jeong_2003}. Here, the offset by 1 represents the finite chance of a publication currently without citations to get cited for the first time. The temporal term represents forgetting.

To this point, the model covers the bulk of average papers. However, as noted before, there are milestone works that are clearly exceptional. These works are causing visible spikes in the forgetting curve Fig.~\ref{fig:aor2} (b) and thus follow a different attachment kernel. If we assume, that the underlying preferential attachment mechanism is the same also for milestones, we can determine their attachment kernel by measuring the value of the exponent $\alpha_m$ of any individual milestone paper $m$. For our analysis, we define milestones as the top thirty most cited papers in the APS, excluding review papers. For each of these papers, we measure the exponent $\alpha_m$ based on the waiting times between events, where these papers receive citations. For details, see~\hyperref[SI4]{SI 4}. The attachment kernel for milestones has the form:
\begin{equation}
   f(k, t, \alpha_m)= (k+1)(t+1)^{-\alpha_m} \quad,
\end{equation}
where $\alpha_m$ can be different for every individual milestone, $m$.
With these ingredients, we can now define the generative model as follows. An exponentially increasing number of papers is published over time. Each of these papers, $i$, cites multiple other papers $j$ that were published before. The probability of $j$ to be cited by $i$ is proportional to the attachment kernel $f(k, t)$, where $k$ is the number of citations that paper $j$ has at that point in time and $t$ is the time difference in years between the publishing times of paper $i$ and paper $j$. To normalize the probability, $f(k,t)$ is calculated for all average papers and  $f(k,t,\alpha)$ for all milestone papers at each point in time. Following this model, one can generate a simulated citation network.

\section*{Results}

\subsection*{Null model validation} 
\begin{figure}[h!]
	\centering
	  \includegraphics[width=0.7\columnwidth]{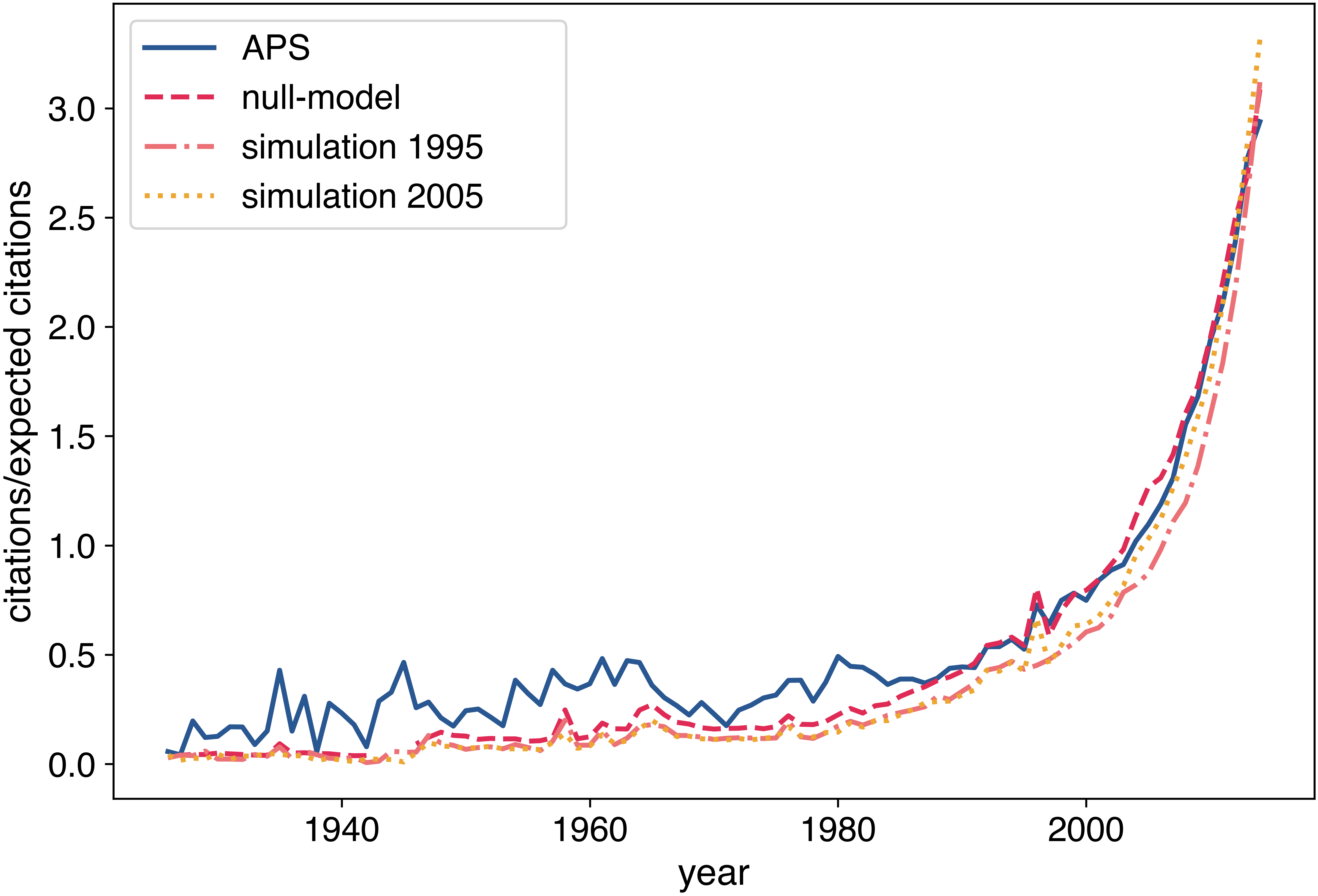}
		\caption{Forgetting curve for the APS citation network for the empirical data (blue), the null model based on attachment kernel $f(k, t)$ (red) and the out-of-sample predictions for the years 1995-2015 (orange) and 2005-2015 (yellow), based on empirical data from 1893 to 1995 and 1893 to 2005 respectively.}
	\label{fig:nullcomparison}
\end{figure}

To test the model, we generate a {\em randomized} citation network that serves as a null model. To generate it, we take the publishing dates and out-degrees of papers from the APS, but omit the information about which paper cites which. The process of citing is replaced by a random process, where probabilities for individual papers to receive a citation are proportional to their respective attachment kernel, $f(k, t)$. We include the publishing dates and individual attachment kernels $f(k, t, \alpha_m)$ for milestones in this model. We then assign the references paper by paper, ordered in time, until we have reconstructed the full citation network. For details, see~\hyperref[SI3]{SI 3}. On the reconstructed network, we measure the forgetting curve. In Fig.~\ref{fig:nullcomparison} the resulting forgetting curve for the null model (red dashed line) is compared to the empirical forgetting curve  (blue solid line) that was introduced in Fig.~\ref{fig:aor2} (b). 

The general features of the null-model are in good agreement with the empirical values, especially for the more recent decades. To assess how far the null model deviates from the empirical data, we calculate the mean absolute percentage error (MAPE) on the forgetting curves. This value is calculated by taking the absolute value of the relative deviation between null model and empirical data for each data point. The resulting values are then averaged to arrive at the MAPE. When comparing the null model to empirical data, the MAPE is below $6\%$ for the last 30 years. It increases for the earlier years, up to values of $35\%$ if the whole 120 years of the APS, including early years with low numbers of publications and high volatility, are taken into account. The low MAPE for recent years implies that the null model accurately reproduces the forgetting curve of the real APS. Note that also some spikes caused by milestone papers are present in the null model. The magnitude of these spikes is captured well for years after the 1990s. The similarity of the null model and the empirical data implies that the attachment kernel was measured accurately and  captures the essential underlying citation dynamics.

\subsection*{Out-of-sample prediction} Next, we estimate the predictive power of the model by asking if the model can be used to simulate plausible future citation networks. To test the predictive power, we split our empirical data in two parts, separated by age, and try to predict the second part using data only from the first part. The quality of the prediction can then be assessed by comparing the prediction results with the empirical data. The first part consists of all papers published until 1995, the second part that we predict are all papers published between 1995 and 2015. We assume (realistically) that the number of references per paper stays more or less constant and that the number of publications published per year grows exponentially with a rate $\gamma = 0.0587$ corresponding to a doubling time of $11.8$ years as determined in \cite{Martin2013}. For our prediction, we use the attachment kernel that we measured on the full APS citation network using data until 1995 only. We simulate the occurrence of new milestone papers with a Poisson process with exponentially distributed waiting times with an estimated rate $\lambda$ from the empirical data. The simulation works as for the null model, but instead of taking the publishing dates and out-degrees from empirical data, we assume an exponentially increasing stream of publications with a constant number of references. We bootstrap the empirical citation network until 1995 and simulate the next 20 years for the forgetting curve of 2015. We then repeat the process on a shorter time period of ten years, between 2005 and 2015.

For the predicted citation network, we compute the forgetting curve and compare it to the empirical one from the APS, see Fig. \ref{fig:nullcomparison}. The orange line shows the forgetting curve of the simulation. It yields a reasonable  approximation to the empirical forgetting curve (blue line). The MAPE for the last 30 years is around $21\%$. This means, that the last three decades of the forgetting curve we predicted for 2015 based on data up to the year 1995 on average deviates  less than $21\%$ from the empirical forgetting curve. The second prediction, for the years between 2005 and 2015, is shown in yellow. Here, the MAPE for the last 30 years is around $17\%$. If all earlier years, back to the 19th century, are included, the error increases up to values of $50\%$. Note that the absence of a spike in the year 1996 in the orange curve is due to the simulation only using publication data until 1995.

\subsection*{Predicting milestones} We next test the predictive power of the model for milestone papers by asking how predictive the exponent $\alpha_m$ is for a paper becoming a milestone in the future. We first define a set of \emph{potential milestones} as all those papers that had at least 200 citations by the end of  2005, a condition that is matched for 814 papers in the data set. Next, {\em true milestones} are defined as those papers that receive at least 50 citations in the year, 10 years after they received their first 200 citations. At this citation rate  they would be ranked among the top cited papers within a few decades. For each potential milestone, $m$, we measure the exponent, $\alpha_m$, based on the first 200 citations.
If this exponent is small, the forgetting can be over-compensated by the preferential attachment mechanism. Thus, if the measured exponent is below a threshold, $\alpha_m \leq \alpha_c$, we predict that the paper will receive more than 50 citations in one year after ten years.

\begin{figure}[h!]
	\centering
	  \includegraphics[width=0.7\columnwidth]{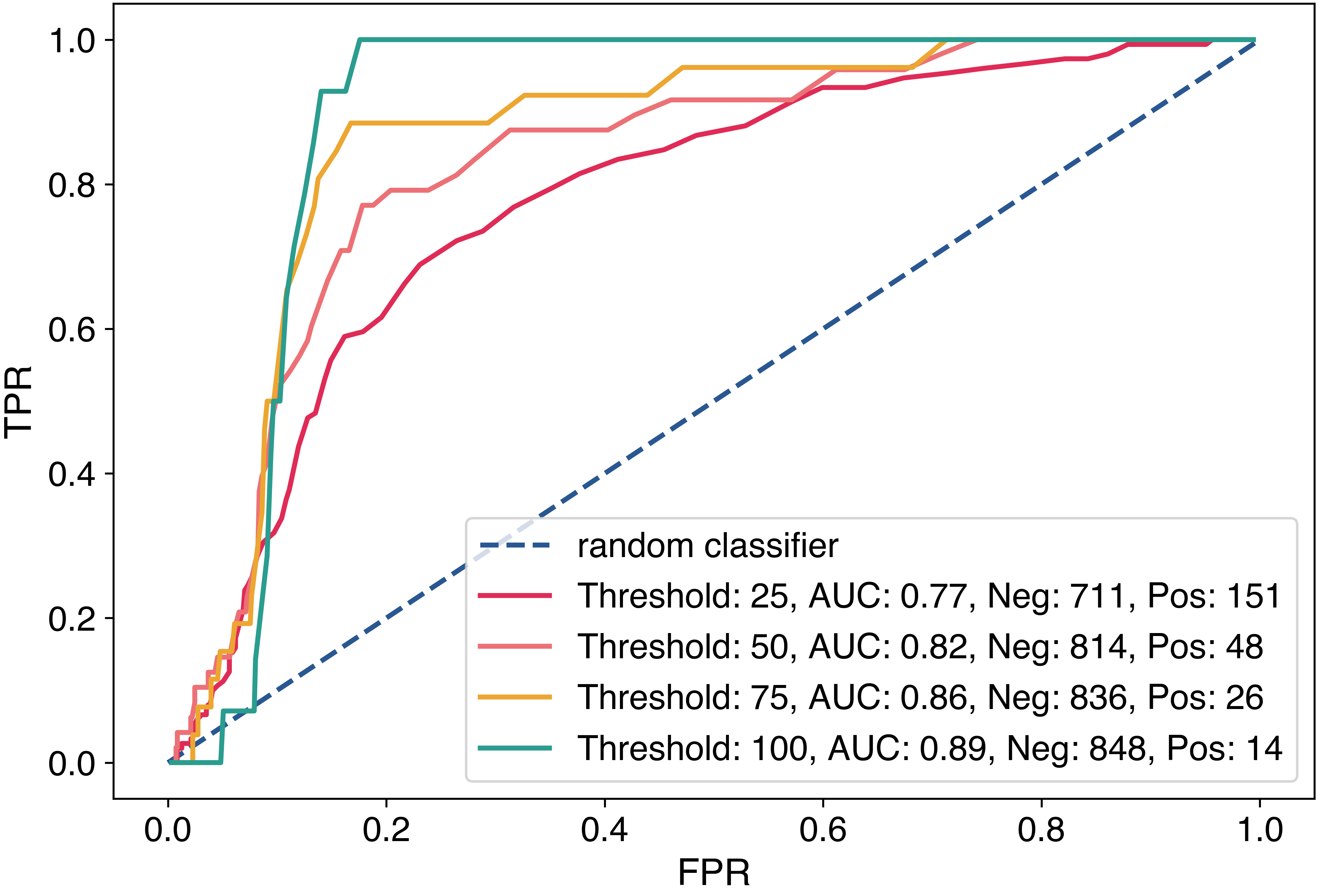}
	  \includegraphics[width=0.7\columnwidth]{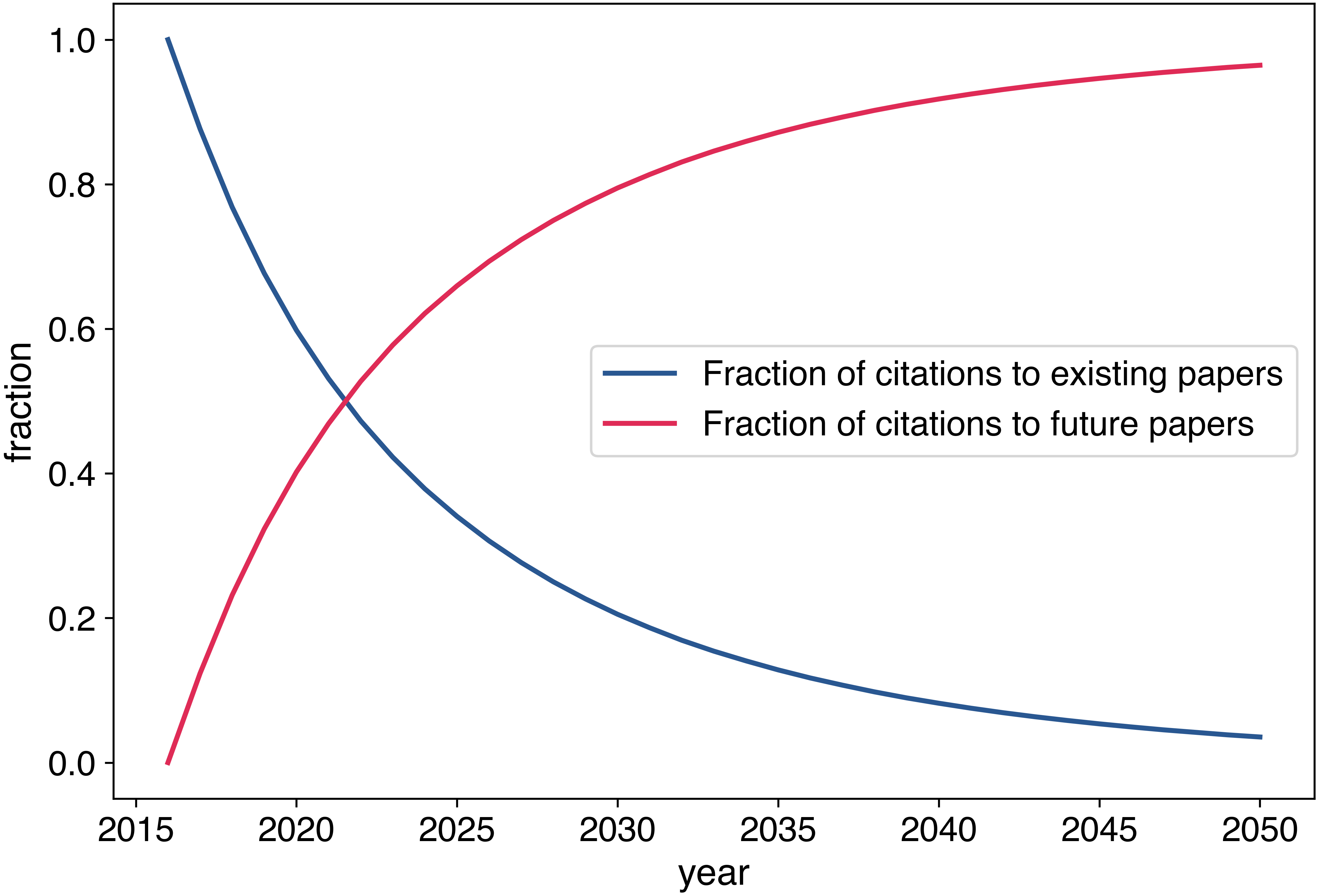}
		\caption{(a) ROC curve (true positive rate vs. false positive rate) for predicting the success of potential milestone papers, $m$, after 10 years, based on their exponent, early in their existence, $\alpha_m$. We show the ROC curves for different thresholds. These thresholds are defined as the minimum number of citations, that a paper must receive in one year, ten years after we measured its exponent $\alpha_m$. The legend shows the values for the area under the curve (AUC) as well as for the number of positives (POS) and negatives (NEG) in our data set for each treshold.
		(b) What we will cite in the future? For every year on the x-axis we consider the references of papers published in that year. The blue curve represents the fraction of references to papers published up to --and including-- 2015. The red curve is the fraction of references to papers, published after 2015. Roughly 20\% of citations from papers published in 2030 are going to papers published until 2015. $96\%$ of the papers cited 30 years from now are not yet written. 
		}
	\label{fig:roc}
\end{figure}

The predictive power of this binary classifier is seen in the ROC curve in Fig. \ref{fig:roc} (a), where we find an AUC of up to 0.89. The high value for the AUC suggests that $\alpha_m$ is a good predictor of the future success of a paper, with high specificity and sensitivity. While our definition of {\em true milestones} was somewhat arbitrary, we confirm that a wide range of thresholds results in reasonably high AUC values and the choice of threshold thus does not significantly impact our results. Our definition of \emph{potential milestones} is based on the requirement for a high number of data points in order to get a low-variance measurement of the exponent $\alpha_m$. For details, see~\hyperref[SI4]{SI 4}.
Note that on the left side of the ROC curve in Fig. \ref{fig:roc} (a) there is a convex section that indicates, that there are some papers that initially have a very small  $\alpha_m$ exponent but fail to stay popular in the long run and thus are erroneously classified as milestones in our model. We hypothesize that the convex section is caused by papers with a sudden but brief success -- ''flash in the pan'' papers.   

For the prediction, we estimate an individual constant $\alpha_m$ based on the first 200 citations of each potential milestone. If instead of taking the first 200 citations, these exponents $\alpha_m$ are derived from several hundreds of citations, a closer look reveals that many of the exponents come with a positive slope as a function of the number of citations $k$. This indicates that for milestone papers, the attachment kernel cannot be factorized as $f(k,t, \alpha_m)=g(k) h(t, \alpha_m)$. For that reason, if one is interested in predicting citation trajectories of individual papers with more than a few hundred citations, $\alpha_m$ should not be taken as a constant but rather as a function of $k$. We find, that this function is best approximated by a linear function $\alpha_m(k) = \alpha_{m_{0}} + \beta k$, which can be estimated from the data in an analogous way. For details, see~\hyperref[SI2]{SI 2}.

\subsection*{Predicting future citation landscapes} After confirming that the model accurately predicts the general shape of the forgetting curve, as well as the success of individual papers, we use it to predict what citations today's research will get. We simulate the future citation network as in the situation of the out-of-sample prediction. We bootstrap it on all empirical data until the end of 2015 and then simulate the next 35 years into the future, based on a continued exponential increase of papers and on the attachment kernel measured until 2015. We simulate the network by introducing paper by paper, until we arrive at the possible citation network for the year 2050.

From this simulated citation network for the years 2016 to 2050, we estimate how quickly today's work will be forgotten. For each of these years, we define the fraction of citations that go to publications published before --and including-- 2015, which we call \textit{existing papers}. The fraction of citations that go to publications published after 2015 we call \textit{future papers}. Figure \ref{fig:roc} (b) shows that the fraction of citations to \textit{existing papers} decreases to about $30\%$ within ten years into the future and reaches a value of $3.6\%$ in 2050. If we shift the curve by six years to estimate what will happen to papers published in 2021, we conclude that more than $95\%$ of the papers cited in 2050 are not yet written. 

\subsection*{Empirical citation trends}
The empirical data of the APS reveals additional trends that are important in our context. First, the average number of citations that papers published in a certain year receive, has been decreasing since the 1960s. And second, larger and larger fractions of our references consist of comparably old and highly cited milestone papers, while the fraction of milestones among the newly published papers has been steadily declining. For details, see~\hyperref[SI6]{SI 6}.

\section*{Discussion}

In this paper, we focused on understanding how scientific work is being utilized or forgotten. In empirical citation networks, we observe two main mechanisms: a linear preferential attachment and a power-law temporal forgetting. We capture these empirical findings in a generative model in which newly published papers cite older papers with a probability proportional to an attachment kernel $f(k, t)$. The kernel reflects two main factors: the cited paper's academic age and the number of citations it already received. This result implies that an article's probability of receiving a citation grows linearly with the citations already received and decays with its publication age as a power-law, characterised by an exponent $\alpha$.

While there is a strong preference towards citing recent papers, some milestone works --potentially of exceptional impact-- continue to attract citations even after decades. These singular milestones, $m$, are characterized by an exponent $\alpha_m < 0.85$. In comparison, the exponent for the bulk of non-exceptionally successful papers is about  $\alpha \sim 1.4 $. Based on $\alpha_m$, one can predict whether a paper will continue to be successful beyond the age of ten years. The corresponding ROC analysis yields an AUC of $0.83$, confirming that the exponent is a good predictor indeed.

%\textbf{Discussion} 
The model reproduces the highly skewed distributions of the number of citations and the age of references observed in other works. In particular, the forgetting curve that results from the model shows a strong tendency to cite recent papers~\cite{deSollaPrice1965, Burton1960}, while older papers fade progressively in oblivion~\cite{deSollaPrice1965, Redner2005, Burton1960}. The spikes observed in the forgetting curve highlight the impact of milestone papers that are remembered for a long time~\cite{Redner2005, Golosovsky2017}.

The linear preferential attachment mechanism in our model confirms the linear dependence on the number of citations, originally found in ~\cite{Jeong_2003}. The second mechanism we find is power-law-like forgetting. While forgetting, or aging, is a mechanism found in most models of citation dynamics, its detailed functional description remained controversial~\cite{Yin2017, Borner2004, Dorogovtsev2000, Valverde2007, Higham2017_2, Safdari2016, Golosovsky2017}. In~\cite{Borner2004} an author-paper network yields an attachment kernel with an aging component that follows a Weibull distribution. \cite{Golosovsky2017, Higham2017_2} both investigate the citation rate of individual papers and find exponential aging, while~\cite{Valverde2007} studies a patent citation network and finds power-law aging with an exponential cutoff in {\em intrinsic time}, where each new patent represents one timestep~\cite{Safdari2016, Dorogovtsev2000},  both find power-law aging in intrinsic time. \cite{Dorogovtsev2000} applies it to the context of citation networks while~\cite{Safdari2016} applies it to a Hollywood-actor network. Our model finds a power-law-like forgetting mechanism with an exponent larger than one for the bulk of average papers. For milestone scientific papers, we find values below one. Power-law-like forgetting implies that papers are forgotten relatively quickly and only have a limited time to attract attention. However, power laws also indicate that old papers keep a chance to get cited. This result enables a few notable publications that attracted much attention early on to continue attracting citations for a long time. 
%In the model presented here, we use real times, not the intrinsic times, to make it independent of the rate at which papers are being published.

%\textbf{Limitations} 
There are several limitations to our model. In particular, it relies on the assumption that the way scientific literature evolves will not change fundamentally. This includes the assumption that the scientific output continues to grow exponentially, an assumption that might not hold forever~\cite{Brown2011}. This observation limits the prediction of the number of citations a paper will receive in the future, simply because the total number of citations is directly proportional to the number of articles published. However, our results only partially depend on this assumption, as the general shape of the forgetting curve is also recovered well for sub-exponential growth and even for linear growth. On the other hand, our model also assumes no fundamental changes in how citations are chosen. This assumption might not hold if, for instance, disruptive new technologies are introduced and adopted globally, as had been the case with the internet.

To forecast citation trajectories of individual highly cited papers, our model provides a simple indicator based only on the citation network.
For higher accuracy, additional parameters, such as the popularity of the authors, the size of the field, or the novelty of the publication, should be taken into account~\cite{ST2020, Klimek2016, Newman2009x, Martin2013, Newman_2014x, Acuna2012, Shen2014, Sinatra2016, Wang2013}. However, even taking these factors into account, there remains uncertainty due to unpredictable events such as a highly relevant discovery that went unnoticed for a while, termed "sleeping beauty" ~\cite{vanRaan2004} or the rediscovery of a paper outside its original domain, called exaptation~\cite{ferreira2020}.

Restricting our model to only age and number of citations makes it easily transferable to other domains. The possiblity that the life-cycle of scientific literature might be different in physics than for other fields like biology, medicine, or philosophy, and even in non-scientific domains such as the music and entertainment industry, is an interesting research direction.

%\textbf{Final remarks} 

Our scheme allows for projecting the current citation ecosystem into the future and investigating its sustainability. Through the empirical forgetting kernel, and under the assumption that the number of papers will keep growing exponentially, it is possible to simulate possible future citation networks. For instance, one learns that 95\% of today's (and yesterday's) work will be outdated before the end of the next three decades. It is this finding, and the fact that papers are receiving fewer and fewer citations on average and have decreasing chances of becoming a milestone, that has substantial implications both for individuals and for scientific policies. Partially, the observed dynamics might be attributed to an increasing fragmentation in the system, driven by the emergence of sub-fields. The fragmentation becomes evident, if one looks at the increase in the number of journals in the APS over the last decades. In this fragmented system, papers might become isolated inside of sub-fields and reach smaller scientific audiences on average. On the other hand, a much more worrisome interpretation of our findings hints towards an increasing quantity-over-quality dynamic, where increasingly short lived publications receive little attention and have diminishing chances of becoming works of great impact.

\section*{Material and Methods}

\subsection*{APS citation data}
In our empirical analysis, we study the citation network of the American Physical Society (APS). This data set consists of 577870 publications that were published between the years 1893 and 2015 in 17 different journals (PR, PRA, PRAB, PRAPPLIED, PRB, PRC, PRD, PRE, PRFLUIDS, PRI, PRL, PRMATERIALS, PRPER, PRSTAB, PRSTPER, PRX, RMP). It includes 6713993 citation pairs between these papers.

\subsection*{Citation Networks}

There are different types of citation networks. In its simplest form, a citation network, $M_{ij}$, captures which paper $j$ cites paper $i$. If paper $j$ cites paper $i$ then $M_{ij} = 1$, otherwise $M_{ij} = 0$. These networks should not be confused with other popular citation based networks such as the co-citation network or the bibliographic coupling network. The co-citation network, $C_{ij} = MM^T$, contains the information of how often papers $i$ and $j$ appear {\em together} in the reference list of another paper \cite{Small1973}. $C_{ij}=2$ means that there exist two papers that cite both, $i$ and $j$. The bibliographic coupling network, $B_{ij} = M^TM$, on the other hand, states how many references papers $i$ and $j$ have in common \cite{MARTYN1964}. Every paper, $i$, is uniquely associated with a publication date, $t_i$, which makes it possible to bring in the time component of citations.

\bibliography{main}

\begin{thebibliography}{10}
\urlstyle{rm}
\expandafter\ifx\csname url\endcsname\relax
  \def\url#1{\texttt{#1}}\fi
\expandafter\ifx\csname urlprefix\endcsname\relax\def\urlprefix{URL }\fi
\expandafter\ifx\csname doiprefix\endcsname\relax\def\doiprefix{DOI: }\fi
\providecommand{\bibinfo}[2]{#2}
\providecommand{\eprint}[2][]{\url{#2}}

\bibitem{Sinatra2015x}
\bibinfo{author}{Sinatra, R.}, \bibinfo{author}{Deville, P.},
  \bibinfo{author}{Szell, M.}, \bibinfo{author}{Wang, D.} \&
  \bibinfo{author}{Barab{\'{a}}si, A.-L.}
\newblock \bibinfo{journal}{\bibinfo{title}{A century of physics}}.
\newblock {\emph{\JournalTitle{Nature Physics}}} \textbf{\bibinfo{volume}{11}},
  \bibinfo{pages}{791--796}, \doiprefix\url{10.1038/nphys3494}
  (\bibinfo{year}{2015}).

\bibitem{Martin2013}
\bibinfo{author}{Martin, T.}, \bibinfo{author}{Ball, B.},
  \bibinfo{author}{Karrer, B.} \& \bibinfo{author}{Newman, M. E.~J.}
\newblock \bibinfo{journal}{\bibinfo{title}{Coauthorship and citation patterns
  in the physical review}}.
\newblock {\emph{\JournalTitle{Physical Review E}}}
  \textbf{\bibinfo{volume}{88}}, \doiprefix\url{10.1103/physreve.88.012814}
  (\bibinfo{year}{2013}).

\bibitem{NSF}
\bibinfo{title}{{National Science Foundation} publications output: U.s. trends
  and international comparisons}.
\newblock
  \bibinfo{howpublished}{\url{https://ncses.nsf.gov/pubs/nsb20206/publication-output-by-region-country-or-economy}}
  (\bibinfo{year}{2021}).
\newblock \bibinfo{note}{Accessed: 2021-02-09}.

\bibitem{Meho_2007}
\bibinfo{author}{Meho, L.~I.}
\newblock \bibinfo{journal}{\bibinfo{title}{The rise and rise of citation
  analysis}}.
\newblock {\emph{\JournalTitle{Physics World}}} \textbf{\bibinfo{volume}{20}},
  \bibinfo{pages}{32--36}, \doiprefix\url{10.1088/2058-7058/20/1/33}
  (\bibinfo{year}{2007}).

\bibitem{Loken2017}
\bibinfo{author}{Loken, E.} \& \bibinfo{author}{Gelman, A.}
\newblock \bibinfo{journal}{\bibinfo{title}{Measurement error and the
  replication crisis}}.
\newblock {\emph{\JournalTitle{Science}}} \textbf{\bibinfo{volume}{355}},
  \bibinfo{pages}{584--585}, \doiprefix\url{10.1126/science.aal3618}
  (\bibinfo{year}{2017}).

\bibitem{Ioannidis2005}
\bibinfo{author}{Ioannidis, J. P.~A.}
\newblock \bibinfo{journal}{\bibinfo{title}{Why most published research
  findings are false}}.
\newblock {\emph{\JournalTitle{{PLoS} Medicine}}} \textbf{\bibinfo{volume}{2}},
  \bibinfo{pages}{e124}, \doiprefix\url{10.1371/journal.pmed.0020124}
  (\bibinfo{year}{2005}).

\bibitem{Redner2005}
\bibinfo{author}{Redner, S.}
\newblock \bibinfo{journal}{\bibinfo{title}{Citation statistics from 110 years
  {ofPhysical} review}}.
\newblock {\emph{\JournalTitle{Physics Today}}} \textbf{\bibinfo{volume}{58}},
  \bibinfo{pages}{49--54}, \doiprefix\url{10.1063/1.1996475}
  (\bibinfo{year}{2005}).

\bibitem{deSollaPrice1965}
\bibinfo{author}{de~Solla~Price, D.~J.}
\newblock \bibinfo{journal}{\bibinfo{title}{Networks of scientific papers}}.
\newblock {\emph{\JournalTitle{Science}}} \textbf{\bibinfo{volume}{149}},
  \bibinfo{pages}{510--515}, \doiprefix\url{10.1126/science.149.3683.510}
  (\bibinfo{year}{1965}).

\bibitem{seglen1992}
\bibinfo{author}{Seglen, P.~O.}
\newblock \bibinfo{journal}{\bibinfo{title}{The skewness of science}}.
\newblock {\emph{\JournalTitle{Journal of the American Society for Information
  Science}}} \textbf{\bibinfo{volume}{43}}, \bibinfo{pages}{628--638}
  (\bibinfo{year}{1992}).

\bibitem{Redner1998}
\bibinfo{author}{Redner, S.}
\newblock \bibinfo{journal}{\bibinfo{title}{How popular is your paper? an
  empirical study of the citation distribution}}.
\newblock {\emph{\JournalTitle{The European Physical Journal B}}}
  \textbf{\bibinfo{volume}{4}}, \bibinfo{pages}{131--134},
  \doiprefix\url{10.1007/s100510050359} (\bibinfo{year}{1998}).

\bibitem{Price1976}
\bibinfo{author}{Price, D. D.~S.}
\newblock \bibinfo{journal}{\bibinfo{title}{A general theory of bibliometric
  and other cumulative advantage processes}}.
\newblock {\emph{\JournalTitle{Journal of the American Society for Information
  Science}}} \textbf{\bibinfo{volume}{27}}, \bibinfo{pages}{292--306},
  \doiprefix\url{https://doi.org/10.1002/asi.4630270505}
  (\bibinfo{year}{1976}).
\newblock
  \eprint{https://asistdl.onlinelibrary.wiley.com/doi/pdf/10.1002/asi.4630270505}.

\bibitem{Barabsi1999x}
\bibinfo{author}{Barab{\'{a}}si, A.-L.} \& \bibinfo{author}{Albert, R.}
\newblock \bibinfo{journal}{\bibinfo{title}{Emergence of scaling in random
  networks}}.
\newblock {\emph{\JournalTitle{Science}}} \textbf{\bibinfo{volume}{286}},
  \bibinfo{pages}{509--512}, \doiprefix\url{10.1126/science.286.5439.509}
  (\bibinfo{year}{1999}).

\bibitem{Krapivsky2001}
\bibinfo{author}{Krapivsky, P.~L.} \& \bibinfo{author}{Redner, S.}
\newblock \bibinfo{journal}{\bibinfo{title}{Organization of growing random
  networks}}.
\newblock {\emph{\JournalTitle{Physical Review E}}}
  \textbf{\bibinfo{volume}{63}}, \doiprefix\url{10.1103/physreve.63.066123}
  (\bibinfo{year}{2001}).

\bibitem{Dorogovtsev2002}
\bibinfo{author}{Dorogovtsev, S.~N.} \& \bibinfo{author}{Mendes, J. F.~F.}
\newblock \bibinfo{journal}{\bibinfo{title}{Evolution of networks}}.
\newblock {\emph{\JournalTitle{Advances in Physics}}}
  \textbf{\bibinfo{volume}{51}}, \bibinfo{pages}{1079--1187},
  \doiprefix\url{10.1080/00018730110112519} (\bibinfo{year}{2002}).

\bibitem{Albert2002}
\bibinfo{author}{Albert, R.} \& \bibinfo{author}{Barab{\'{a}}si, A.-L.}
\newblock \bibinfo{journal}{\bibinfo{title}{Statistical mechanics of complex
  networks}}.
\newblock {\emph{\JournalTitle{Reviews of Modern Physics}}}
  \textbf{\bibinfo{volume}{74}}, \bibinfo{pages}{47--97},
  \doiprefix\url{10.1103/revmodphys.74.47} (\bibinfo{year}{2002}).

\bibitem{Newman2003-x}
\bibinfo{author}{Newman, M. E.~J.}
\newblock \bibinfo{journal}{\bibinfo{title}{The structure and function of
  complex networks}}.
\newblock {\emph{\JournalTitle{{SIAM} Review}}} \textbf{\bibinfo{volume}{45}},
  \bibinfo{pages}{167--256}, \doiprefix\url{10.1137/s003614450342480}
  (\bibinfo{year}{2003}).

\bibitem{Newman2001}
\bibinfo{author}{Newman, M.~E.}
\newblock \bibinfo{journal}{\bibinfo{title}{{Clustering and preferential
  attachment in growing networks}}}.
\newblock {\emph{\JournalTitle{Physical Review E - Statistical Physics,
  Plasmas, Fluids, and Related Interdisciplinary Topics}}}
  \textbf{\bibinfo{volume}{64}}, \bibinfo{pages}{4},
  \doiprefix\url{10.1103/PhysRevE.64.025102} (\bibinfo{year}{2001}).
\newblock \eprint{0104209}.

\bibitem{Capocci2006}
\bibinfo{author}{Capocci, A.} \emph{et~al.}
\newblock \bibinfo{journal}{\bibinfo{title}{Preferential attachment in the
  growth of social networks: The internet encyclopedia wikipedia}}.
\newblock {\emph{\JournalTitle{Physical Review E}}}
  \textbf{\bibinfo{volume}{74}}, \doiprefix\url{10.1103/physreve.74.036116}
  (\bibinfo{year}{2006}).

\bibitem{Jeong_2003}
\bibinfo{author}{Jeong, H.}, \bibinfo{author}{N{\'{e}}da, Z.} \&
  \bibinfo{author}{Barab{\'{a}}si, A.~L.}
\newblock \bibinfo{journal}{\bibinfo{title}{Measuring preferential attachment
  in evolving networks}}.
\newblock {\emph{\JournalTitle{Europhysics Letters ({EPL})}}}
  \textbf{\bibinfo{volume}{61}}, \bibinfo{pages}{567--572},
  \doiprefix\url{10.1209/epl/i2003-00166-9} (\bibinfo{year}{2003}).

\bibitem{Borner2004}
\bibinfo{author}{B{\"{o}}rner, K.}, \bibinfo{author}{Maru, J.~T.} \&
  \bibinfo{author}{Goldstone, R.~L.}
\newblock \bibinfo{journal}{\bibinfo{title}{{The simultaneous evolution of
  author and paper networks}}}.
\newblock {\emph{\JournalTitle{Proceedings of the National Academy of Sciences
  of the United States of America}}} \textbf{\bibinfo{volume}{101}},
  \bibinfo{pages}{5266--5273}, \doiprefix\url{10.1073/pnas.0307625100}
  (\bibinfo{year}{2004}).

\bibitem{Lehmann2005}
\bibinfo{author}{Lehmann, S.}, \bibinfo{author}{Jackson, A.~D.} \&
  \bibinfo{author}{Lautrup, B.}
\newblock \bibinfo{journal}{\bibinfo{title}{Life, death and preferential
  attachment}}.
\newblock {\emph{\JournalTitle{Europhysics Letters ({EPL})}}}
  \textbf{\bibinfo{volume}{69}}, \bibinfo{pages}{298--303},
  \doiprefix\url{10.1209/epl/i2004-10331-2} (\bibinfo{year}{2005}).

\bibitem{Newman2009x}
\bibinfo{author}{Newman, M. E.~J.}
\newblock \bibinfo{journal}{\bibinfo{title}{The first-mover advantage in
  scientific publication}}.
\newblock {\emph{\JournalTitle{{EPL} (Europhysics Letters)}}}
  \textbf{\bibinfo{volume}{86}}, \bibinfo{pages}{68001},
  \doiprefix\url{10.1209/0295-5075/86/68001} (\bibinfo{year}{2009}).

\bibitem{Newman_2014x}
\bibinfo{author}{Newman, M. E.~J.}
\newblock \bibinfo{journal}{\bibinfo{title}{Prediction of highly cited
  papers}}.
\newblock {\emph{\JournalTitle{{EPL} (Europhysics Letters)}}}
  \textbf{\bibinfo{volume}{105}}, \bibinfo{pages}{28002},
  \doiprefix\url{10.1209/0295-5075/105/28002} (\bibinfo{year}{2014}).

\bibitem{Burton1960}
\bibinfo{author}{Burton, R.~E.} \& \bibinfo{author}{Kebler, R.~W.}
\newblock \bibinfo{journal}{\bibinfo{title}{The
  {\textquotedblleft}half-life{\textquotedblright} of some scientific and
  technical literatures}}.
\newblock {\emph{\JournalTitle{American Documentation}}}
  \textbf{\bibinfo{volume}{11}}, \bibinfo{pages}{18--22},
  \doiprefix\url{10.1002/asi.5090110105} (\bibinfo{year}{1960}).

\bibitem{Fanelli2016}
\bibinfo{author}{Fanelli, D.} \& \bibinfo{author}{Larivi{\`{e}}re, V.}
\newblock \bibinfo{journal}{\bibinfo{title}{Researchers' individual publication
  rate has not increased in a century}}.
\newblock {\emph{\JournalTitle{{PLOS} {ONE}}}} \textbf{\bibinfo{volume}{11}},
  \bibinfo{pages}{e0149504}, \doiprefix\url{10.1371/journal.pone.0149504}
  (\bibinfo{year}{2016}).

\bibitem{Wang2013x}
\bibinfo{author}{Wang, D.}, \bibinfo{author}{Song, C.} \&
  \bibinfo{author}{Barab{\'{a}}si, A.-L.}
\newblock \bibinfo{journal}{\bibinfo{title}{Quantifying long-term scientific
  impact}}.
\newblock {\emph{\JournalTitle{Science}}} \textbf{\bibinfo{volume}{342}},
  \bibinfo{pages}{127--132}, \doiprefix\url{10.1126/science.1237825}
  (\bibinfo{year}{2013}).

\bibitem{Yin2017}
\bibinfo{author}{Yin, Y.} \& \bibinfo{author}{Wang, D.}
\newblock \bibinfo{journal}{\bibinfo{title}{{The time dimension of science:
  Connecting the past to the future}}}.
\newblock {\emph{\JournalTitle{Journal of Informetrics}}}
  \textbf{\bibinfo{volume}{11}}, \bibinfo{pages}{608--621},
  \doiprefix\url{10.1016/j.joi.2017.04.002} (\bibinfo{year}{2017}).
\newblock \eprint{1704.04657}.

\bibitem{Dorogovtsev2000}
\bibinfo{author}{Dorogovtsev, S.~N.} \& \bibinfo{author}{Mendes, J.~F.}
\newblock \bibinfo{journal}{\bibinfo{title}{{Evolution of networks with aging
  of sites}}}.
\newblock {\emph{\JournalTitle{Physical Review E - Statistical Physics,
  Plasmas, Fluids, and Related Interdisciplinary Topics}}}
  \textbf{\bibinfo{volume}{62}}, \bibinfo{pages}{1842--1845},
  \doiprefix\url{10.1103/PhysRevE.62.1842} (\bibinfo{year}{2000}).

\bibitem{Safdari2016}
\bibinfo{author}{Safdari, H.} \emph{et~al.}
\newblock \bibinfo{journal}{\bibinfo{title}{{Fractional dynamics of network
  growth constrained by aging node interactions}}}.
\newblock {\emph{\JournalTitle{PLoS ONE}}} \textbf{\bibinfo{volume}{11}},
  \bibinfo{pages}{1--13}, \doiprefix\url{10.1371/journal.pone.0154983}
  (\bibinfo{year}{2016}).

\bibitem{Golosovsky2017}
\bibinfo{author}{Golosovsky, M.} \& \bibinfo{author}{Solomon, S.}
\newblock \bibinfo{journal}{\bibinfo{title}{{Growing complex network of
  citations of scientific papers: Modeling and measurements}}}.
\newblock {\emph{\JournalTitle{Physical Review E}}}
  \textbf{\bibinfo{volume}{95}}, \bibinfo{pages}{1--19},
  \doiprefix\url{10.1103/PhysRevE.95.012324} (\bibinfo{year}{2017}).

\bibitem{Higham2017}
\bibinfo{author}{Higham, K.~W.}, \bibinfo{author}{Governale, M.},
  \bibinfo{author}{Jaffe, A.~B.} \& \bibinfo{author}{Z{\"{u}}licke, U.}
\newblock \bibinfo{journal}{\bibinfo{title}{{Fame and obsolescence:
  Disentangling growth and aging dynamics of patent citations}}}.
\newblock {\emph{\JournalTitle{Physical Review E}}}
  \textbf{\bibinfo{volume}{95}}, \bibinfo{pages}{1--7},
  \doiprefix\url{10.1103/PhysRevE.95.042309} (\bibinfo{year}{2017}).
\newblock \eprint{1611.05076}.

\bibitem{Higham2017_2}
\bibinfo{author}{Higham, K.~W.}, \bibinfo{author}{Governale, M.},
  \bibinfo{author}{Jaffe, A.~B.} \& \bibinfo{author}{Z{\"{u}}licke, U.}
\newblock \bibinfo{journal}{\bibinfo{title}{{Unraveling the dynamics of growth,
  aging and inflation for citations to scientific articles from specific
  research fields}}}.
\newblock {\emph{\JournalTitle{Journal of Informetrics}}}
  \textbf{\bibinfo{volume}{11}}, \bibinfo{pages}{1190--1200},
  \doiprefix\url{10.1016/j.joi.2017.10.004} (\bibinfo{year}{2017}).
\newblock \eprint{1708.08335}.

\bibitem{Valverde2007}
\bibinfo{author}{Valverde, S.}, \bibinfo{author}{Sol{\'{e}}, R.~V.},
  \bibinfo{author}{Bedau, M.~A.} \& \bibinfo{author}{Packard, N.}
\newblock \bibinfo{journal}{\bibinfo{title}{{Topology and evolution of
  technology innovation networks}}}.
\newblock {\emph{\JournalTitle{Physical Review E - Statistical, Nonlinear, and
  Soft Matter Physics}}} \textbf{\bibinfo{volume}{76}}, \bibinfo{pages}{1--7},
  \doiprefix\url{10.1103/PhysRevE.76.056118} (\bibinfo{year}{2007}).
\newblock \eprint{0612030}.

\bibitem{Brown2011}
\bibinfo{author}{Brown, J.~H.} \emph{et~al.}
\newblock \bibinfo{journal}{\bibinfo{title}{Energetic limits to economic
  growth}}.
\newblock {\emph{\JournalTitle{{BioScience}}}} \textbf{\bibinfo{volume}{61}},
  \bibinfo{pages}{19--26}, \doiprefix\url{10.1525/bio.2011.61.1.7}
  (\bibinfo{year}{2011}).

\bibitem{ST2020}
\bibinfo{author}{Thurner, S.}, \bibinfo{author}{Liu, W.},
  \bibinfo{author}{Klimek, P.} \& \bibinfo{author}{Cheong, S.~A.}
\newblock \bibinfo{journal}{\bibinfo{title}{The role of mainstreamness and
  interdisciplinarity for the relevance of scientific papers}}.
\newblock {\emph{\JournalTitle{{PLOS} {ONE}}}} \textbf{\bibinfo{volume}{15}},
  \bibinfo{pages}{e0230325}, \doiprefix\url{10.1371/journal.pone.0230325}
  (\bibinfo{year}{2020}).

\bibitem{Klimek2016}
\bibinfo{author}{Klimek, P.}, \bibinfo{author}{{S. Jovanovic}, A.},
  \bibinfo{author}{Egloff, R.} \& \bibinfo{author}{Schneider, R.}
\newblock \bibinfo{journal}{\bibinfo{title}{{Successful fish go with the flow:
  citation impact prediction based on centrality measures for term–document
  networks}}}.
\newblock {\emph{\JournalTitle{Scientometrics}}}
  \textbf{\bibinfo{volume}{107}}, \bibinfo{pages}{1265--1282},
  \doiprefix\url{10.1007/s11192-016-1926-1} (\bibinfo{year}{2016}).

\bibitem{Acuna2012}
\bibinfo{author}{Acuna, D.~E.}, \bibinfo{author}{Allesina, S.} \&
  \bibinfo{author}{Kording, K.~P.}
\newblock \bibinfo{journal}{\bibinfo{title}{Predicting scientific success}}.
\newblock {\emph{\JournalTitle{Nature}}} \textbf{\bibinfo{volume}{489}},
  \bibinfo{pages}{201--202}, \doiprefix\url{10.1038/489201a}
  (\bibinfo{year}{2012}).

\bibitem{Shen2014}
\bibinfo{author}{Shen, H.}, \bibinfo{author}{Wang, D.}, \bibinfo{author}{Song,
  C.} \& \bibinfo{author}{Barab\'{a}si, A.-L.}
\newblock \bibinfo{title}{Modeling and predicting popularity dynamics via
  reinforced poisson processes}.
\newblock In \emph{\bibinfo{booktitle}{Proceedings of the Twenty-Eighth AAAI
  Conference on Artificial Intelligence}}, AAAI'14, \bibinfo{pages}{291–297}
  (\bibinfo{publisher}{AAAI Press}, \bibinfo{year}{2014}).

\bibitem{Sinatra2016}
\bibinfo{author}{Sinatra, R.}, \bibinfo{author}{Wang, D.},
  \bibinfo{author}{Deville, P.}, \bibinfo{author}{Song, C.} \&
  \bibinfo{author}{Barabasi, A.-L.}
\newblock \bibinfo{journal}{\bibinfo{title}{Quantifying the evolution of
  individual scientific impact}}.
\newblock {\emph{\JournalTitle{Science}}} \textbf{\bibinfo{volume}{354}},
  \bibinfo{pages}{aaf5239--aaf5239}, \doiprefix\url{10.1126/science.aaf5239}
  (\bibinfo{year}{2016}).

\bibitem{Wang2013}
\bibinfo{author}{Wang, D.}, \bibinfo{author}{Song, C.} \&
  \bibinfo{author}{Barab{\'{a}}si, A.-L.}
\newblock \bibinfo{journal}{\bibinfo{title}{Quantifying long-term scientific
  impact}}.
\newblock {\emph{\JournalTitle{Science}}} \textbf{\bibinfo{volume}{342}},
  \bibinfo{pages}{127--132}, \doiprefix\url{10.1126/science.1237825}
  (\bibinfo{year}{2013}).

\bibitem{vanRaan2004}
\bibinfo{author}{van Raan, A. F.~J.}
\newblock \bibinfo{journal}{\bibinfo{title}{Sleeping beauties in science}}.
\newblock {\emph{\JournalTitle{Scientometrics}}} \textbf{\bibinfo{volume}{59}},
  \bibinfo{pages}{467--472}, \doiprefix\url{10.1023/b:scie.0000018543.82441.f1}
  (\bibinfo{year}{2004}).

\bibitem{ferreira2020}
\bibinfo{author}{Ferreira, M.~R.} \emph{et~al.}
\newblock \emph{\bibinfo{title}{Quantifying Exaptation in Scientific
  Evolution}}, \bibinfo{pages}{55--68} (\bibinfo{publisher}{Springer
  International Publishing}, \bibinfo{year}{2020}).

\bibitem{Small1973}
\bibinfo{author}{Small, H.}
\newblock \bibinfo{journal}{\bibinfo{title}{Co-citation in the scientific
  literature: A new measure of the relationship between two documents}}.
\newblock {\emph{\JournalTitle{Journal of the American Society for Information
  Science}}} \textbf{\bibinfo{volume}{24}}, \bibinfo{pages}{265--269},
  \doiprefix\url{10.1002/asi.4630240406} (\bibinfo{year}{1973}).

\bibitem{MARTYN1964}
\bibinfo{author}{Martyn, J.}
\newblock \bibinfo{journal}{\bibinfo{title}{{Bibliographiy} {Coupling}}}.
\newblock {\emph{\JournalTitle{Journal of Documentation}}}
  \textbf{\bibinfo{volume}{20}}, \bibinfo{pages}{236--236},
  \doiprefix\url{10.1108/eb026352} (\bibinfo{year}{1964}).

\end{thebibliography}

\section*{Author contributions statement}
NR, ST, VS, VL conceptionalized the work and designed the model. NR, MF and WS did the data preparation. NR and WS did the implementation. NR did the analysis. NR and ST produced a first draft, all contributed to the final paper.

\section*{Additional information}
\textbf{Competing interests}\\
The authors declare no conflict of interest.

% \begin{figure}[ht]
% \centering
% \includegraphics[width=\linewidth]{stream}
% \caption{Legend (350 words max). Example legend text.}
% \label{fig:stream}
% \end{figure}

\newpage

\section*{Supplementary information}
\subsection*{Text S1: Milestone significance}
\label{SI1}

\begin{figure}[h!]
	\centering
	 \includegraphics[width=0.7\columnwidth]{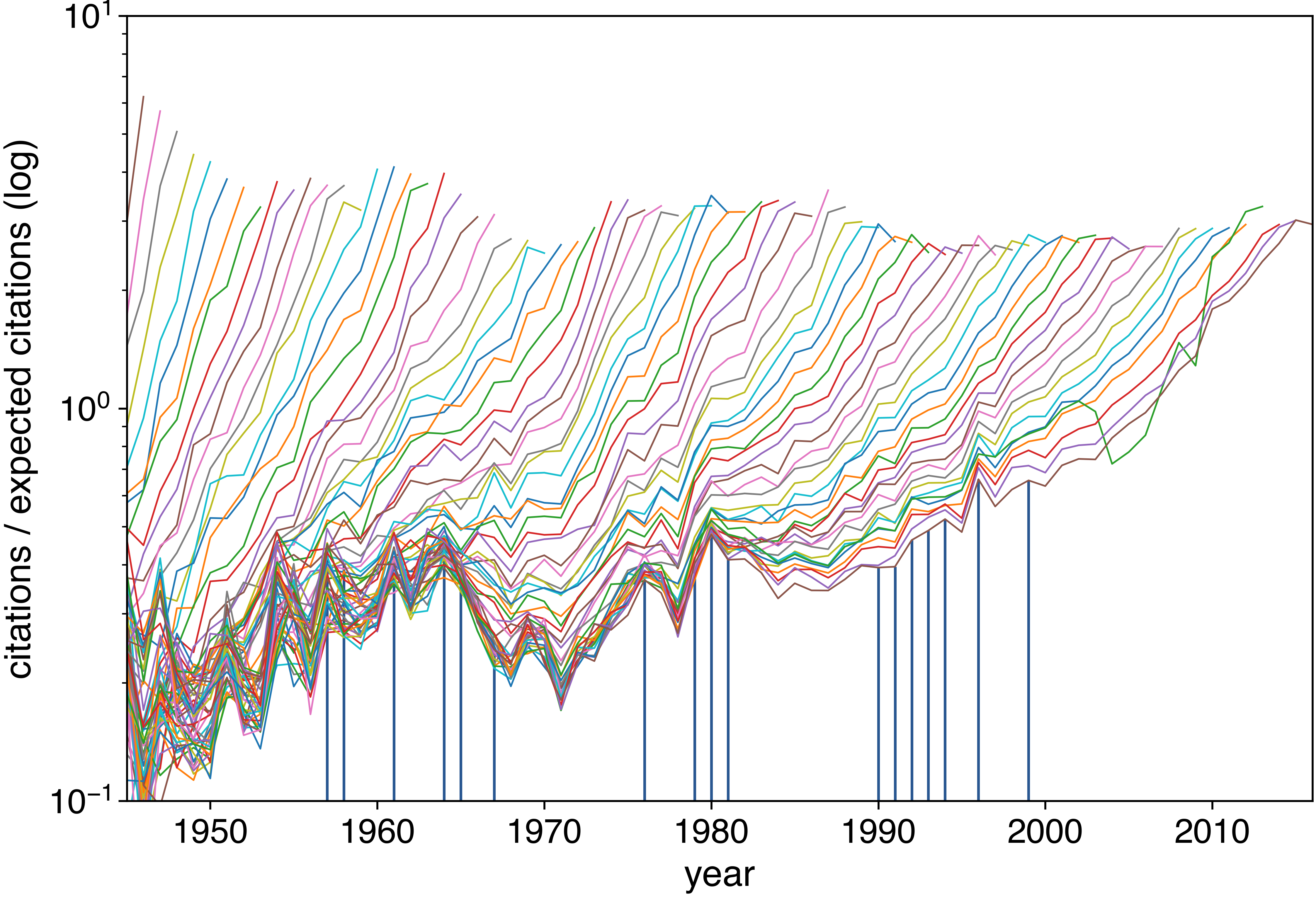}
	\caption{The forgetting-curves of the APS. Each line represents the forgetting-curve for papers published in the year after the right-most point of the curve (e.g. if the curve ends in 2014, it is the forgetting curve of papers published in 2015). Years where milestone papers were published are marked with vertical lines. In these years, one can observe the formation of spikes due to the adoption of the milestone paper by following the vertical line from the top (earlier years) to the bottom (recent years).
	}
	\label{fig:vitosplot}
\end{figure}

In the forgetting curve in Fig.~\ref{fig:aor2} (b) we observed spikes coinciding with the publishing years of the top 30 papers. SI Fig.~\ref{fig:vitosplot} shows how these spikes develop over the years.

We want to show that these spikes can be caused by works of exceptional quality, here called milestones. To do so, we calculate the relative share $cit_{ms}$ of citations that go to milestone papers, compared to citations that go to all papers. The calculation is simple: For each year, we count the number of citations to milestone papers published in that year, divided by the total number of citations to all papers published in that year.

We then average this value over all the years when milestone papers were published. We find, that in the years when milestone papers were published, they attract around $6\%$ of all citations. If we repeat the analysis but consider only citations from papers published in 2015, the last year of our data set, we find that in the years when milestone papers were published, these milestones attract around $14\%$ of all citations of papers published in 2015. Since the forgetting curve is directly proportional to the number of citations, these values are indeed high enough to cause a visible spike in the forgetting curve.

\subsection*{Text S2: Measuring the attachment kernel}
\label{SI2}

\begin{figure}[h!]
	\centering
	 \includegraphics[width=0.9\columnwidth]{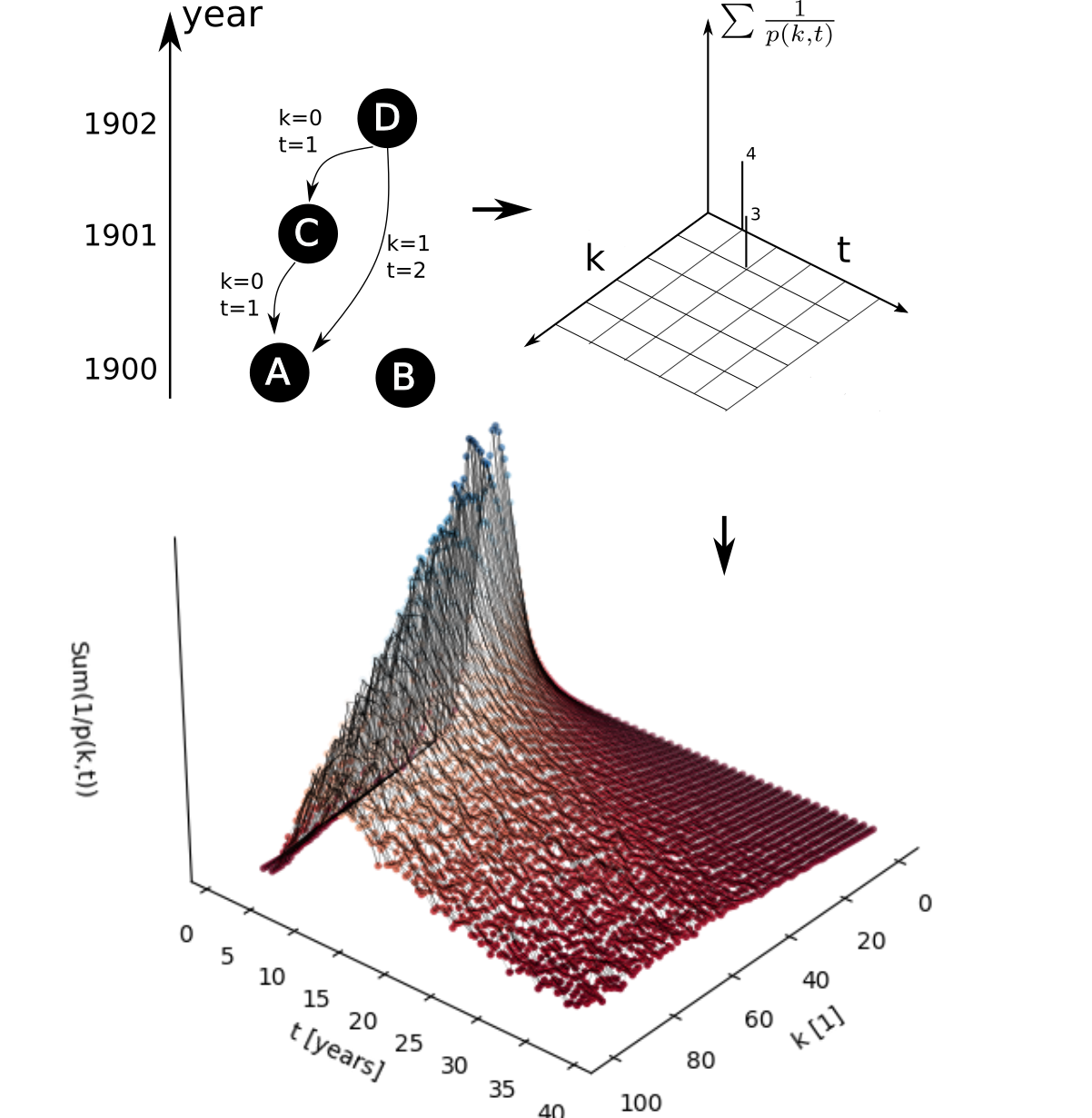}
	\caption{Approximating the kernel $f(k, t)$ of the APS citation network.\\
	a) The citation network is rebuilt paper by paper. Each back dot is a paper, each link is a citation. The papers are published in alphabetical order. For reach citation, the values of k, t and the state of the network are used to build the histogram.\\
	b) The histogram is generated as the sum of the inverse probabilities when reconstructing the citation network. \\
	c) The histogram for the entire network. This landscape in k and t represents the shape of the probability kernel $f(k, t)$. The shape is only representative for areas where sufficient data is available.
	}
	\label{fig_newman}
\end{figure}

To measure the attachment kernel, $f(k,t)$, we modify the method originally used by Newman in co-authorship networks \cite{Newman2001} to two dimensions. We then apply it to the 120 years of empirical citation data of the American Physical Society.
In this method, we go through the papers of the APS, one by one, ordered in time, and build a histogram from which we can read off the attachment kernel. The steps included are as follows. First, we go through the APS citation network $M_{i,j}$ node by node and link by link,  starting from the first paper ever published and ending on the most recent paper. At each point in time, the probability of a paper $j$ citing a specific paper $i$ with age,  $t$, and $k$ citations, is given by:
\begin{equation}
    P(k,t) = f(k, t) \frac{n(t_j)}{N(t_j)} \, , 
\end{equation}
where $n(t_j)$ is the number of papers present at publishing time $t_j$ of paper $j$ that have the same exact values of $k$ and $t$ as paper $i$,  and $N(t_j)$ is the total number of papers present at that time. Then the attachment kernel $f(k,t)$ can be estimated from a histogram where each contribution is weighted with the inverse probability $\frac{N(t_j)}{n(t_j)}$. A sketch of this method is shown in Fig. \ref{fig_newman}.
From the resulting histogram, we can extract slices for fixed values of $k$ or $t$. These let us fit the functions in $k$ and $t$ respectively. If the slices only differ in a factor, then the kernel is of the form $f(k,t)=g(k)h(t)$. We find, that this is true in good approximation for the bulk of average publications. 

The result is a two-dimensional histogram that approximates the probability kernel, $f(k, t)$. Note, however, that many combinations of $(k, t)$ are underrepresented in the history of the APS and thus also in the histogram. For example, it is evident that there are very few papers that have a high number of citations already in the first few years. Thus, the histogram is only representative in areas where sufficient data is available. For this reason, it is not practicable to directly fit a function to the histogram.

By fixing $k$ or $t$ to specific values, slices in $t$ or $k$ can be extracted from the surface. We slice the landscape for all integer values of $k$ and $t$ and fit functions to the slices. If the probability kernel is factorizable, $f(k, t)=g(k)h(t)$, then the slices would only differ by a factor. We fit a linear function $g(k) = k$ and a power $h(t) = t^{-\alpha}$. From this, we calculate an average exponent as the weighted average over all exponents, where the weights correspond to the number of times a particular value of $k$ was observed in the history of the APS.

\subsection*{Text S3: Milestone attachment kernel}
\label{SI4}

In the introduction, we discussed that some old publications receive exceptionally high numbers of citations. We call these the milestone papers. In the context of a citation network, we define them through their citations and take the top 30 most cited papers of all time that are not review papers. The choice is arbitrary, but the results hold for different values as long as the papers are receiving numerous citations. 

The milestone papers are exceptions, they stand out from the bulk. As such, their number is low, and the statistics are not sufficient to measure their attachment kernel with the method described in~\hyperref[SI2]{SI 2}. Instead, we propose a slightly different method, described below. 

To figure out an estimator for the probability of being cited, we draw a comparison to an urn model. Suppose we have an urn with different colors of balls. We are only interested in one special color. There are multiple copies of that color in the urn. In each run, we draw with replacement until we draw a ball of our color. If we do many runs, and on average it takes $n$ draws until a ball of our color is drawn, then the most plausible estimator for the probability of drawing our color would be $\frac{1}{n}$.

In addition to that, we introduce the concept of intrinsic time for a citation network. Each citation event, where any paper is cited, represents one discrete time step in intrinsic time. For a milestone paper m, the probability of being cited at any intrinsic time step $\tau$ is given by:
\begin{equation}
    p_m = \frac{f(k_{m, \tau}, t_m, \alpha_m)}{\sum_i f(k_{i, \tau}, t_i, \alpha_i)} \, ,
\end{equation}
where f is the probability kernel of the form:
\begin{equation}
    f_i = f(k_{i, \tau}, t_i, \alpha_i) = (k_{i, \tau}+1)(t_i + 1)^{-\alpha_i} \, , 
\end{equation}
and $k_{i, \tau}$ is the in-degree of paper $i$ at intrinsic time-step $\tau$, $t_i$ is the age of paper $i$ in years and $\alpha_i$ is the forgetting-exponent of paper $i$.\\
The most likely estimator for the number of citations $C$ it takes until paper m is cited is $\frac{1}{p_m}$. We can estimate:
\begin{equation}
    C \approx \frac{1}{p_m} = \frac{\sum_i f_{i, \tau}}{(k_{m, \tau}+1)(t_m + 1)^{-\alpha_m}}
\end{equation}
leading to 
\begin{equation}
\label{eq_alpha}
    \alpha_m \approx \log(\frac{C (k_{m, \tau}+1)}{\sum_i f_{i, \tau}}) \frac{1}{\log(t_{m, \tau}+1)} \, .
\end{equation}
We know $k_{m, \tau}$ and  $t_{m}$. $C$ is the number of citation events between two events where the milestone paper $m$ is cited. 
The $\alpha_i$ are unknown, but they only appear in the sum. Since we have a good approximation for them, supposedly, the sum should be approximated correctly. Thus, we can estimate $\alpha_m$ directly. For each time the milestone paper $m$ is cited, we get a different $C$ and $k_{m, \tau}$ is increased by 1, so we collect a set of values for $\alpha_m$
\begin{equation}
\label{eq:msalpha}
    \overline{\alpha}_m = \log(\frac{C(k_m+1)}{\sum_i f_{i, \tau}})\frac{1}{\log(t_{m, \tau}+1)} \, . 
\end{equation}
These values can be averaged (before applying the logarithm) in order to determine an average exponent. However, if we apply a linear fit to the set of values for $\alpha$, we notice a general slightly increasing trend. This suggests that in our approximation for the kernel, $f(k,t) = (k+1)(t + 1)^{-\alpha}$, the functions of $k$ and $t$ cannot be separated completely for milestone papers. Instead, there is a small dependency on $k$ in $\alpha$. A more fitting approximation for milestone papers is thus given by $f(k,t) = (k+1)(t + 1)^{-(a+bk)}$ i.e., the exponent $\alpha$ is a linear function of $k$ with a very small and positive slope.

\begin{figure}[h!]
	\centering
	 \includegraphics[width=0.7\columnwidth]{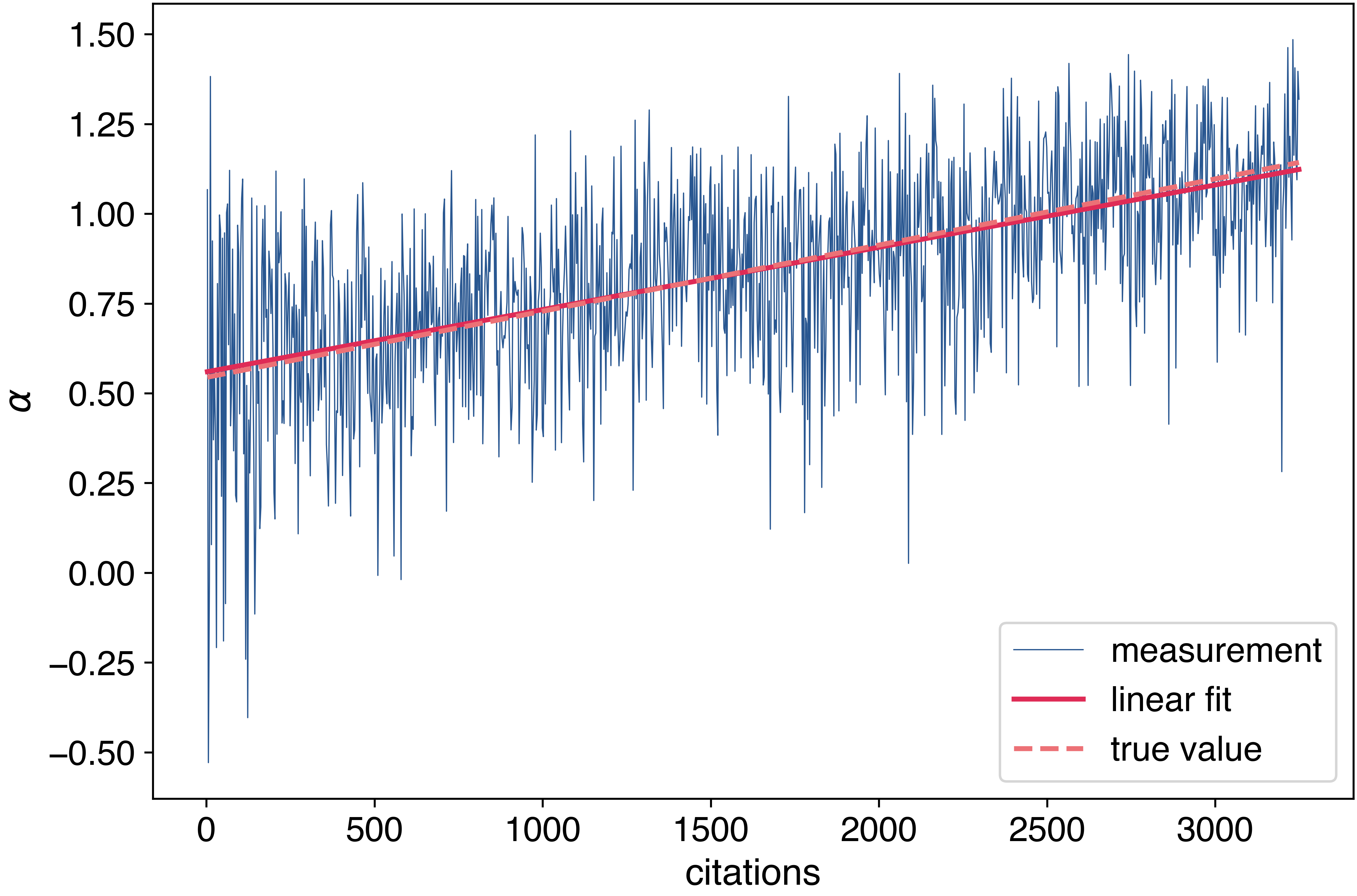}
	\caption{Exponent $\alpha$ of $(k+1)(t + 1)^{-\alpha}$ vs in-degree k for a representative milestone paper, measured on the simulated null model. Blue: individual measurements of $\alpha$, Red: linear fit on the individual measurements, Orange, dashed: true value (simulation input)
	}
	\label{fig_mspexp}
\end{figure}

To validate the method, we construct the null model explained in~\hyperref[SI3]{SI 3}. We simulate the full system and then repeat the measurement of the exponent in the simulated system. Since the exponent in the attachment kernel of the simulation are configuration parameters, we have perfect knowledge of them. Thus, we can compare the values we measure with the ones we used as an input. If the measurement method is valid, we would expect these values to be the same, apart from some noise. Indeed, we measure an average absolute deviation of $\Delta\alpha=0.1$. If binning is used to reduce the noise, the average absolute deviation can be reduced to $\Delta\alpha=0.01$. The result for one example paper is shown in fig. \ref{fig_mspexp}.

Two additional effects might have an impact on the measured value and thus need to be ruled out. First, in equation \ref{eq:msalpha} the sum $\sum_i f_{i,\tau}$ changes after each citation event (each intrinsic time step $\tau$). For the calculation, we take the value at the end of the interval, i.e., the value of $\tau$ at the time when the milestone is cited. This is technically inaccurate, but the inaccuracy is very small. We measure this both at the beginning and the end of the interval, confirming that the sum changes only by an insignificant amount. Thus, this effect can be neglected.
Second, each paper cites n other papers, but each of the cited papers can only appear once in the reference list. Thus, the probabilities for getting cited by one paper change depending on what else was cited by the same paper. Also, this effect can be neglected since on average the number of references per paper is very small compared to the total number of citable papers. 

\subsection*{Text S4: Null model}
\label{SI3}

\begin{figure}[h!]
	\centering
	 \includegraphics[width=0.7\columnwidth]{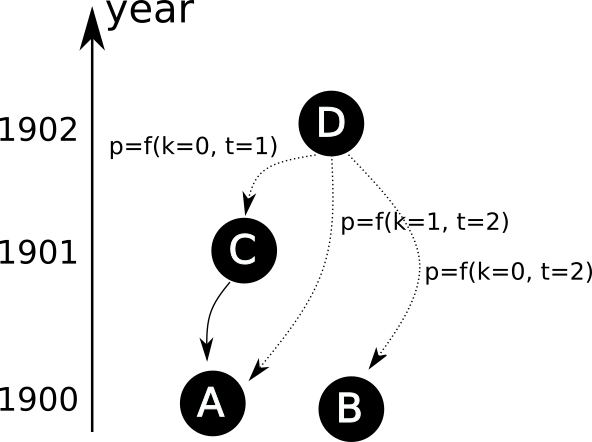}
	\caption{Null model of the APS citation dynamics. Papers are published in the same order, at the same time and with the same out-degree (number of references) as the papers of the APS citation network. Each paper randomly cites papers that were published before with a probability proportional to the probability kernel, $f(k,t)$.
	}
	\label{fig:null}
\end{figure}

We construct a null-model of the APS citation dynamics. In this model, we keep all publishing dates and out-degrees (number of references) of real papers from the empirical data, but we attach each outgoing link to a random paper that was published before based on the probability kernel $f(k,t)$. For milestone papers we use the exponent that we determined from the real data, for the bulk of average papers we take the average exponent. A sketch of the null-model is shown in  SI Fig. \ref{fig:null}.

\subsection*{Text S5: Citation trends}
\label{SI6}

\begin{figure}[h!]
	\centering
	 \includegraphics[width=0.7\columnwidth]{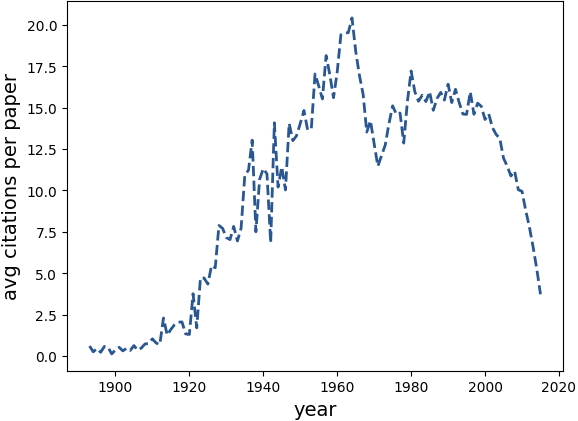}
	\caption{
	Average number of citations that papers published in each year between 1893 and 2015 received. Note that the steep decline during the last decade is due to the papers not having had enough time to accumulate all their citations yet.
	}
	\label{fig_SI_cit}
\end{figure}

To reveal general citation trends in the APS, we investigate the average number of citations that a paper published in a certain year receives, see fig. \ref{fig_SI_cit}. Note that the steep decline during the last decade is due to the papers not having had enough time to accumulate all their citations yet.

\begin{figure}[h!]
	\centering
	 \includegraphics[width=0.7\columnwidth]{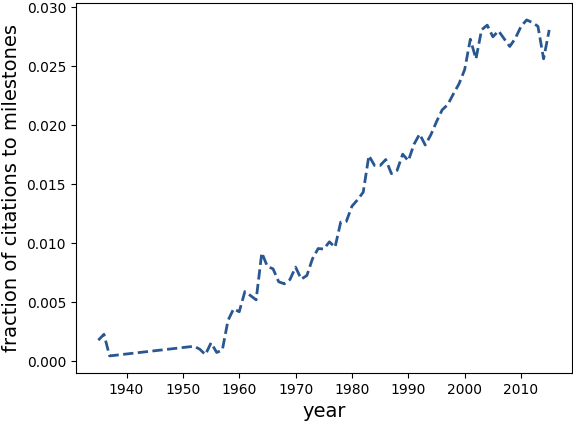}
	\caption{
	    Fraction of references to highly cited papers ($\geq 1000$ citations) over references to all papers for each year between 1893 and 2015.
	}
	\label{fig_SI_ms1}
\end{figure}

To show that an increasing number of citations go to milestone papers, we plot the fraction of references of all papers published in a certain year, that reference highly cited papers (papers with $\geq 1000$ citations) over the total number of references. This is shown in fig. \ref{fig_SI_ms1}.  

\begin{figure}[h!]
	\centering
	 \includegraphics[width=0.7\columnwidth]{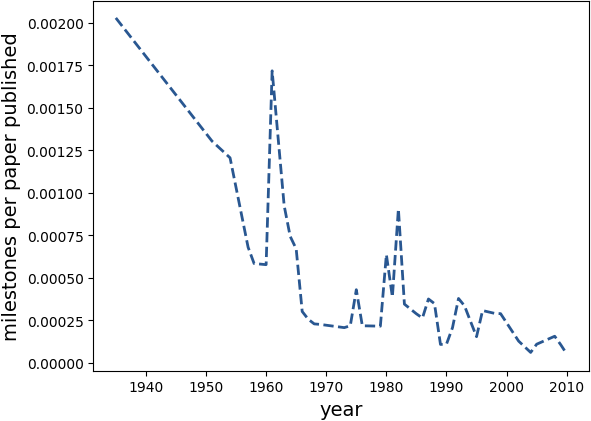}
	\caption{
	Fraction of highly cited papers published ($\geq 1000$ citations) over total number of papers published for each year between 1893 and 2015.
	}
	\label{fig_SI_ms2}
\end{figure}

In fig. \ref{fig_SI_ms2} we show the number of highly cited papers (papers with $\geq 1000$ citations) published in a certain year over the total number of papers published in that year.

\end{document}